\def\cm{{\rm\thinspace cm}}

\def\erg{{\rm\thinspace erg}}

\def\K{{\rm\thinspace K}}
\def\keV{{\rm\thinspace keV}}

\def\kpc{{\rm\thinspace kpc}}

\def\Msun{\hbox{$\rm\thinspace M_{\odot}$}}

\def\s{{\rm\thinspace s}}
\def\yr{{\rm\thinspace yr}}

\def\pcmcuK{\hbox{$\cm^{-3}\K\,$}}

\def\ergps{\hbox{$\erg\s^{-1}\,$}}

\def\Msunpyr{\hbox{$\Msun\yr^{-1}\,$}}

\def\psqcm{\hbox{$\cm^{-2}\,$}}

\documentstyle[psfig,times]{mn}
\begin{document}
\title{On the soft X-ray spectrum of cooling flows}
\author[]
{\parbox[]{6.in} {A.C.~Fabian$^1$, R.F.~Mushotzky$^2$, 
P.E.J.~Nulsen$^3$ and J.R.~Peterson$^4$\\
\footnotesize
1. Institute of Astronomy, Madingley Road, Cambridge CB3 0HA \\
2. NASA/GSFC, Code 662, Greenbelt MD20771, U.S.A. \\
3. Department of Engineering Physics, University of Wollongong,
Wollongong, NSW 2522, Australia \\
4. Columbia Astrophysics Laboratory, 550 W 120th St., New York,
NY10027, U.S.A. \\
}}

\maketitle
\begin{abstract}
Strong evidence for cooling flows has been found in low resolution
X-ray imaging and spectra of many clusters of galaxies. However high
resolution X-ray spectra of several clusters from the Reflection
Grating Spectrometer (RGS) on XMM-Newton now show a soft X-ray
spectrum inconsistent with a simple cooling flow. The main problem is
a lack of the emission lines expected from gas cooling below 1--2~keV.
Lines from gas at about 2--3~keV are observed, even in a high
temperature cluster such as A\,1835, indicating that gas is cooling
down to about 2--3~keV, but is not found at lower temperatures. Here
we discuss several solutions to the problem; heating, mixing,
differential absorption and inhomogeneous metallicity. Continuous or
sporadic heating creates further problems, including the targetting of
the heat at the cooler gas and also the high total energy required. So
far there is no clear observational evidence for widespread heating,
or shocks, in cluster cores, except in radio lobes which occupy only
part of the volume. The implied ages of cooling flows are short, at
about 1~Gyr. Mixing. or absorption, of the cooling gas are other
possibilities. Alternatively, if the metals in the intracluster medium
are not uniformly spread but are clumped, then little line emission is
expected from the gas cooling below 1~keV. The low metallicity part
cools without line emission whereas the strengths of the soft X-ray
lines from the metal-rich gas depend on the mass fraction of that gas
and not on the abundance, since soft X-ray line emission dominates the
cooling function below 2~keV.
\end{abstract}


\section{Introduction}
X-ray spectra of the cores of clusters with cooling flows made with
the Reflection Grating Spectrometer (RGS; den Herder et al 2000) on
XMM-Newton show a remarkable lack of emission lines from gas at 1~keV
or below. For the cluster A\,1835, at redshift $z=0.2523$, both
Chandra (Schmidt et al 2000) and XMM-Newton (Peterson et al 2000) CCD
spectra show a strong temperature decrease towards the cluster centre
from about 9~keV down to about 3~keV. RGS spectra (Peterson et al
2000) are well fitted by a model of gas cooling over that energy
range, at a rate of about $1000\Msunpyr,$ but not to lower
temperatures. There is clear evidence for emission lines from FeXXIV,
but not from lower ionization stages such as FeXX--XXII and in
particular FeXVII which characterize gas cooling through temperatures
of 1~keV and below, respectively. The limit on the rate of gas cooling
below 1~keV is about $150\Msunpyr$. A similar temperature decrease
inward is seen in Chandra (Fabian et al 2000b) and XMM-Newton (Tamura
et al 2000) CCD spectra of A\,1795 ($z=0.063$), again with no emission
lines from gas cooling through and below 1~keV apparent in the RGS
spectra (Tamura et al 2000). The low temperature cooling rate limit in
this case is about $140\Msunpyr$, which is close to the total cooling
rate inferred from ASCA spectra (Fabian et al 1994; Allen et al 1999).

Various explanations for this disagreement with simple cooling flow
expectations (see Fabian 1994 for a review) are given by Peterson et
al (2000). Just invoking a heat source is not sufficient, unless it
has very specific properties, since gas does not appear to accumulate
at 3~keV, it appears to cool down to that temperature then vanish.
Moreover, since gas at 3~keV occupies less than 6 per cent of the
volume of gas at 9~keV, any heating needs to be targetted. It is
possible that the 3~keV gas mixes with hotter or much colder gas, or
absorption becomes important for cooler gas. We discuss these problems
and possibilities here in more detail.

Finally we advance a new possibility involving strong metallicity
variations in the intracluster gas. The line radiation from
intracluster gas above 3 keV is proportional to its metallicity, but
if it cools to lower temperatures where line radiation dominates the
cooling function, the predicted line strengths depend on the mass
fraction of the gas which is metal rich. That fraction may be small.

\section{Heating}

Many authors have suggested that radiative cooling in cluster cores is
offset by some form of heating. Some scenarios invoke conduction with
the hot gas being the heat source (Tucker \& Rosner 1983; Bertschinger
\& Meiksin 1986; Gaetz 1989). Others suggest that activity from the
nucleus of the central galaxy is responsible (Rosner
\& Tucker 1989; B\"ohringer \& Morfill 1989; Loewenstein, Zweibel \&
Begelman 1991; David \& Tucker 1997; Binney \& Tabor 1995; Soker et al
2000; David et al 2000).

We note that the mass of gas at about 3~keV in A\,1835 is consistent
with that expected from radiative cooling of the hotter 9~keV gas. The
radiative cooling time, $t_{\rm c},$ of gas at 3~keV is about 20 per
cent of that at 9~keV (for bremsstrahlung cooling at constant pressure
$t_{\rm c}\propto T^{3/2}$; $t_{\rm c}(3\keV)\sim 1$~Gyr or less in
A\,1835), so in a steady flow the mass of gas below 3~keV is only 20
per cent of that below 9 keV, within the flow. If gas were
accumulating at 3~keV for the lifetime of the flow, then there would
be 5 times more of it than expected from steady cooling alone. This is
inconsistent with the observed mass fractions. If steady heating is
responsible then it has to be some form of `complete' heating which
takes the gas back up to the hotter temperature, since there is no
evidence for significant amounts of gas at other temperatures.

One mechanism for this has been suggested by Norman \& Meiksin (1996).
They propose that the magnetic topology of the gas changes as it cools
down and at some point, say 3~keV, reconnection enables rapid thermal
conduction to occur between the cooler gas and the hotter surrounding
gas. Why this should happen throughout the flow at 3~keV is unclear.
It could be that magnetic fields pin cooling blobs into comoving with
the flow until the density contrast becomes so great that they fall
and mix into the surrounding hotter gas, or perhaps this is when
magnetic fields dominate the pressure (for a cooling spherical cloud,
magnetic pressure $P_{\rm B}\propto T^{-4/3}$). As noted by Norman
\& Meiksin (1996), mechanisms such as this do not actually stop the
cooling flow altogether since energy is being lost, unless the hotter
phase is thermally coupled by conduction to the outer hot gas beyond
the cooling flow. Chandra images of 'cold fronts' in some clusters
(Markevitch et al 2000; Vikhlinin et al 2000) indicate that conduction
there is highly suppressed (Ettori \& Fabian 2000). 

Note too that the volume occupied by gas below temperature $T$ in a
cooling flow $\propto T^{5/2}$ if bremsstrahlung dominates;
below a few keV where line cooling dominates then the dependence is
steeper. This means that gas below 3~keV in a flow originating at
9~keV occupies less than 6 per cent of the volume of the hotter gas.
Heating must be targetted at this small volume if cooling down to
3~keV is to take place. The volume is yet smaller if we allow for
gravitational work done on the gas.

Thermal conduction also has a steep temperature ($T^{5/2}$ if
proportional to the Spitzer rate) which means that it will operate
principally in the hotter gas rather than in the cooler gas in the
system (see also discussion by Bregman \& David 1988).

An alternative to steady heating is sporadic heating, The nucleus, for
example, may have short intervals of strong activity interspersed with
periods of relative quiescence. Several heating models have been
proposed along these lines (Rosner \& Tucker 1989; Tucker \& David
1997; Binney \& Tabor 1995, Soker et al 2000). In an analysis of
Chandra data of the Hydra A cluster, David et al (2000) find from
spectra that the rate at which gas cools to low temperatures in the
inner 30~kpc is inconsistent with the rate of gas flow into this
region and invoke sporadic heating by the radio source to heat the
remainder.

Problems with this scenario are a) the large energy input required and
b) efficient coupling of that energy to the gas. For example, the
thermal energy of the gas within 100~kpc of the centre of A\,1835 is
about $3\times10^{61}\erg$. To heat two-thirds of this (i.e. to take
the gas back from $3\keV\rightarrow9\keV$) requires a power of
$2\times 10^{46}\ergps$ for $3\times 10^7\yr$ if the coupling
efficiency is 100 per cent. This exceeds the total power of an extreme
(and rare) radio source such as 3C295 (Allen et al 2000, where the
radio emission in assumed to be only 10 per cent of the radio power).
Most of the power of such sources is however deposited beyond the core
since the lobes propagate through the intracluster medium rapidly
($\sim 10^7\yr$ to cross 100~kpc for 3C295, using the radio hot spot
speed of Perley \& Taylor 1991).

Lower powers could be used for longer times, but then the probability
of seeing a cooling flow being heated becomes significant and no such
example has yet been identified. Observations of cooling flows around
radio sources, e.g. 3C295 (Allen et al 2000) and 3C84 (Fabian et al
2000a), show no evidence for heating taking place beyond the radio
lobes, which occupy only a small fraction of the volume. Indeed in the
case of 3C84 in the Perseus cluster the data show that the {\it
coolest} gas in the core lies immediately around the radio lobes
(Fabian et al 2000a; see also B\"ohringer et al 1995 for M87, where
the gas close to the radio lobes is cooler than the surrounding gas).
Nevertheless, heating of some of the gas (at least that in the lobes)
in cluster cores by radio sources must take place. Whether it can be
sufficient to offset cooling in a massive system such as A\,1835 is
unknown. A major problem is that no cluster yet shows the heating
taking place in a widespread manner; no strong shocks, which Soker et
al (2000) invoke as the main distributed heat source set up by radio
jets, have been found in cluster cores. If heating is the solution
then jets may be considerably more powerful than their radio
luminosity implies.

If a cluster is heated from near to its centre, the details of the
heating process do not have a major effect on the structure of the
heated gas.  The principal issue is the rate of heating relative to
the size of the region that is heated appreciably.  In a heating event
where the heating rate exceeds the thermal energy of the gas divided
by its sound crossing time, a shock forms and the heated gas will
usually be left in a convectively unstable state.  The hottest gas may
then rise well outside the heated core as the atmosphere returns to
convective equilibrium in a few free-fall times.  Lower heating rates
do not produce a shock, although they still tend to make the heated
gas convectively unstable.  However, for low heating rates convective
motion has time to maintain near convective equilibrium while the gas
is being heated, so that the whole of the region that is heated
significantly becomes nearly isentropic.  This means that in order for
slow heating to prevent deposition of cold gas from a cooling flow,
the whole of the region that would otherwise be the cooling flow
becomes isentropic.  This is not consistent with the data if the
temperature declines towards the centres of these clusters is due to
radiative cooling.  Fast heating would also result in a nearly
isentropic core between heating events, but need not make the whole of
the region where there is net heating isentropic (David et al. 2000).
Observed abundance gradients (Ezawa et al 1997) are also inconsistent
with a convective core.

We have previously argued that wind power from accreting black holes
may have been responsible for heating the intracluster medium in all
clusters, in order to explain the observed luminosity--temperature
relation (Wu, Fabian \& Nulsen 2000). We envisage this process
happening at earlier times (redshifts greater than one) than the
heating in the present discussion, and also to originate from most of
the cluster galaxies. Heating by the nucleus of the central cluster
galaxy could be seen as the remnant of a continuum of heating activity
by galactic nuclei, provided the problems discussed above can be
overcome.

A merger with another cluster may also disrupt a cooling flow (see
discussion in Allen et al 1999 and Gomez et al 2000). For the mean age
of cooling flows to be about 1 Gyr or less, as is required from the
typical cooling time of gas at 2--3~keV at the pressure near the
centre of a cluster, the merger rate must be high. Small clusters or
groups are the only plausible merger partners (no evolution is seen in
the cluster luminosity function within $z=0.3$; Ebeling et al 1997).
Again, how such a process could apparently target the coolest X-ray
emitting gas in a cluster core is unclear.

\section{Mixing}

If cooling does take some gas down to say $10^4\K$ then mixing of
hotter gas with that cold gas could cause it to be undetectable in the
X-ray band. In the case of A\,1835 there are considerable amounts of
gas at $10^3-10^4\K$ within a radius of $20\kpc$ from the centre seen
in optical line emission (Allen 1995; Crawford et al 1999), and over
$10^{11}\Msun$ of molecular gas have been reported (Edge et al 2000).

The mixing process could resemble a mixing layer as discussed by
Begelman \& Fabian (1990). Momentum constrains the relative quantities
of gas at high $T_{\rm h}$ and low $T_{\rm l}$ temperatures which mix
such that the final temperature is approximately $\sqrt(T_{\rm h}
T_{\rm l}).$ This means that gas at $3\times 10^7\K$ mixing with gas at
$3\times 10^3\K$ ends at a temperature of $\sim 3\times 10^5\K$.  Such
gas will cool very rapidly and add to the gas mass at the lower
temperature.

Gas cooling within a blob is likely to be thermally unstable, with the
denser cooler gas falling, shredding and mixing into the surrounding
less dense parts. This too could lead to much of the thermal energy of
a blob being radiated in the EUV, rather than in soft X-rays.

The relevance of cool mixing may be testable by UV observations of the
strong line emission expected from gas at $10^5 - 10^6\K$ and optical
observations of coronal line emission. Mixed gas in A\,1795 would have to be
below $10^{5.5}\K$ in order to be consistent with the results of Yan
\& Cohen (1995). Reprocessing of UV emission by colder gas could
explain some of the optical emission lines from the central regions of
cooling flows (Heckman et al 1989; Crawford et al 2000; Donahue et al
2000). The total power radiated in these lines (which is about 20
times that detected in H$\alpha$, for which the luminosity $L({\rm
H}\alpha)$ ranges from $10^{41} - 10^{43}\ergps$) is similar to that
associated with the gas cooling from about 1~keV. There is not
therefore any problem here with the energetics, indeed there is a good
correspondence with the missing energy from gas cooling below 1~keV
and the total power radiated by the optical nebulosity, although
ionization by massive young stars presumably accounts for some of
$L({\rm H}\alpha)$.

\section{Differential absorption}

The deficit of emission below 1~keV seen with ROSAT (Allen \& Fabian
1997) and ASCA (Fabian et al 1994; Allen et al 1999) has often been
ascribed to absorption internal to the cooling flow. The  models used
have however generally been very simple screens across the whole flow. 
If instead the absorption only operated on the cooler gas, say that
below 3~keV which, as noted above, occupies only a few per cent of the
volume, then the X-ray emission lines from cooler gas could be
rendered undetectable. The required column densities would be in the
range of a few $10^{21}$ to $10^{22}\psqcm$. 

If the flow was not highly multiphase, then much of the cooler gas
would be near the centre which is also where most of the absorbing gas
would need to lie. This gas could either be in sheets and filaments of
cooled gas, or cooled parts of cooling clouds. The implied mass of
absorbing gas is similar to that assumed to be deposited by the
cooling flow over a few Gyr (Allen et al 1999). Again, the optical
nebulosity commonly seen in this region may also be involved.
Observation of patches of absorption, in both soft X-ray lines and
continuum would test this possibility.

Resonance scattering may also be important for soft X-ray resonance
lines. The 15.01~A emission of FeXVII, which is produced by gas at
around $4\times 10^6\K$ is a key line missing from the observed
spectra. This line is weakened in Solar spectra by resonance
scattering (see e.g. Rugge \& McKenzie 1985). Using the formula
presented by Rugge \& McKenzie (1985), we find that the optical depth
for that line in a large flow such as in A\,1835 could exceed 100.
(The mass of gas around $4\times 10^6\K$ is obtained from the product
of the cooling time at that temperature, $t_{\rm c}\sim 3\times
10^6\yr$, and the mass cooling rate, $\sim 1000\Msunpyr$; the column
density is then obtained using the pressure,
$\sim2\times10^6\pcmcuK$.) Such a high optical depth can increase the
total path length of the resonant line photons by a factor of a 2--3,
making them more prone to absorption by cold gas. Also, if there is a
widespread distribution of cooling gas, then some line emission can be
scattered beyond the 1~arcmin effective field of view of the RGS.

\section{Metallicity variations}

Consider that the intracluster medium is highly inhomogeneous, not
principally in density or temperature but in $Z$, the metallicity.
Although there is little evidence for this, it is not obvious that the
metals ejected from galaxies and supernovae should mix into a
magnetized hot gas very easily. The strong continuous shear in the
disc of a spiral galaxy may mix gas fairly well there, but the
situation is far different in the intracluster medium. We assume that
the scale of the metal-rich clumps is very small ($\ll1\kpc$) so
is spatially unresolved.

Let 90 per cent of the gas have zero metallicity ($Z=0$) and 10 per
cent have 3 times Solar ($Z=3$), i.e. the metallicity is bimodal. The
spectrum of such gas then looks on average just like one with $Z=0.3$.
The iron ions, for example, radiate in the same way wherever they are
located in the optically-thin intracluster medium.

What happens now as the gas cools. i.e. in the central cooling flow?
If the gas starts at $kT=10\keV$ bremsstrahlung cooling dominates in
all the gas. Line cooling only dominates the cooling function below
about 2 keV for gas around Solar metallicity (see B\"ohringer \&
Hensler 1989; the temperatureat which it happens increases with
metallcity). The metal-rich gas cools together with the metal-poor
gas until about 2 keV below which it cools much faster and drops out.
The line emission from the cooling flow is only from the metal-rich
gas which is only one tenth of the total flow.

The line strengths are not 10 times what they would have been for that
gas had it had Z=0.3, since the gas only has so much thermal energy,
most of which comes out from the lines. The line intensity depends on
$\dot M$ rather than $Z$ where line cooling dominates (it does depend
on $Z$ at higher temperatures).  The net effect is that the cooling
lines from the whole flow are much weaker than they would be were the
gas fully mixed.  In summary, when the gas is not cooling it looks
like gas with $Z=0.3$. But when it is radiatively cooling,
particularly below 2 keV, it looks like much lower metallicity gas.

The X-ray emission from such a inhomogeneous-metallicity flow
resembles the sum of a metal-poor one and a metal-rich one. This is
illustrated in Fig.~1. If the gas is simply of two varieties, one with
metallicity 0 and the other with mass fraction $f$ and metallicity
$Z$, then the cooling flow will approximate a standard (mixed) one of
mass cooling rate $\dot M$ down to where the lines dominate cooling in
the metal-rich gas, with metallicity $fZ$ (the lower panel in Fig.~1
shows that the iron K lines are close in relative strength to those
from the mixed gas). Below that temperature the
soft X-ray lines will be appropriate for a flow of only $f \dot M =
Z_{\rm obs} \dot M / Z$, where $Z_{\rm obs}$ is the observed (average)
metallicity (Fig.~1 shows that the iron L lines are much weaker than
those from the mixed gas). 

\begin{figure}
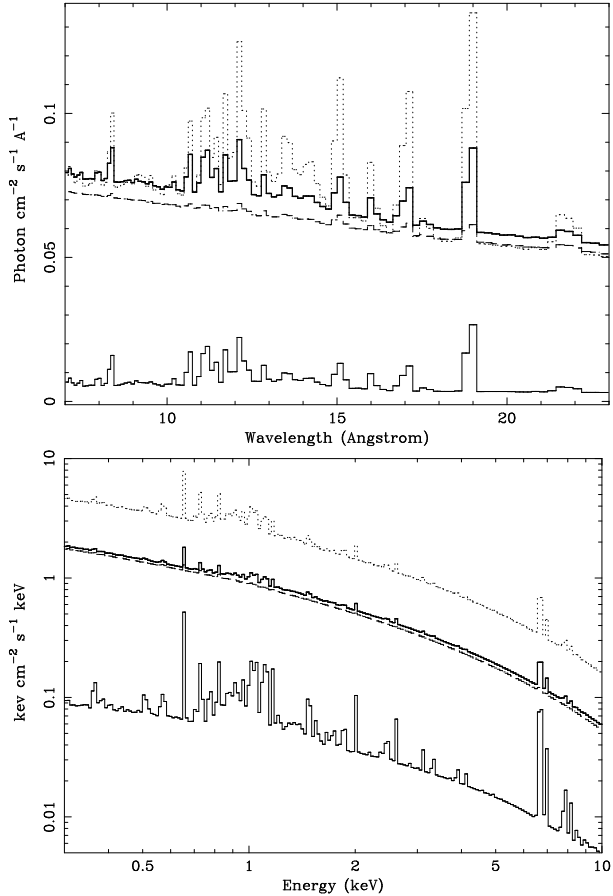

\psfig{figure=rjc_w.ps,width=8truecm,angle=270}
\psfig{figure=rjc_eh.ps,width=8truecm,angle=270}
\caption{Comparison of cooling flow spectra from gas  with differing
metallicities. All have an upper temperature of 9~keV. Upper panel:
the bottom spectrum is $10\Msunpyr$ of gas with $Z=2$, then
successively above are $100\Msunpyr$ at $Z=0.01$, the sum of these two
(bold solid line), and finally $110\Msunpyr$ at $Z=0.2$ (dotted line).
Much of the line emission from 11 -- 17~A is iron L shell emission (XXIV
to XVII roughly going from left to right in the figure, FeXVII
accounts for the emission at 15 and 17~A); OVIII emission is at 19~A.
Lower panel: similar to the upper panel but plotted against energy and
with a range showing the iron K lines; the top curve (dotted line) is
for a $Z=0.2$ flow with a higher rate of $330\Msunpyr$ so there is no
overlap. The spectra of non-cooling hot gases corresponding to the top
two models (a $10:1$ mixture of gas at $Z=0.01$ and $Z=2$, and gas at
$Z=0.2$) would be indistinguishable.}
\end{figure}

Note that if the oxygen abundance is high in a flow (or there were
regions with enhanced oxygen abundance which cool separately) and  they
became an important coolant in the X-ray range then the oxygen line
fluxes depend on the cooling rate, not the abundance too.
Determinations of abundance from cooling flows are inherently
complicated, more so if the metallicities of different elements are
inhomogeneous. This is also relevant to the hot gas commonly seen in
elliptical galaxies.

If the metal-rich gas begins cooling at a larger radius than the
metal-poor gas (the line emission will help to do this, but the
metal-rich gas could also be slightly denser too) then the X-ray
emission spectrum will be complicated to analyse. It is possible that
some of the abundance gradients seen in the cores of some cooling flow
clusters (e.g. Ezawa et al 1997) could be due to these effects.

A possible way to test the inhomogeneous metallicity solution is to
search for metallicity variations in the intracluster medium using
high spatial resolution images (Chandra and XMM-Newton). Also, if the
optical line-emitting nebulosity at the centres of cooling flows is
due to the metal-rich gas then abundance determinations of that gas
may provide a clue.

\section{Discussion}

We have shown that the XMM-Newton RGS soft X-ray spectra of cooling
flows are difficult to interpret. There are problems with many
solutions. Sporadic heating, mixing, differential absorption and
inhomogeneous metallicity offer possible solutions. If the last proves
correct then it has wide implications for the enrichment of both the
intracluster medium, and the intergalactic medium which presumably
would also have a patchy metallicity. Some issues to do with whether
the metals from the first stars are mixed uniformaly into the
intergalactic medium or not are discussed by Rees (1997). The chemical
evolution of galaxies might also become complex, with bimodal
populations\footnote{The globular cluster population of M87 has a
bimodal colour distribution which has been intrepreted as due to the
later merger of a metal-rich galaxy (Kundu et al 1999). Formation from
gas with bimodal metallicity is an alternative possibility.}.

One possible advantage with a heating explanation is that there is no,
or little, mass deposition. This means that the problem of the fate of
the cooled gas in a standard cooling flow is avoided. As already
mentioned however there is strong evidence for massive star formation
and cooled gas in many cooling flows. 

If hot mixing or heating solutions apply to cooling flows then they
may well apply to the formation of all galaxies by cooling. There has
long been an assumption that the visible matter in galaxies is that
which could cool on an infall time (Rees \& Ostriker 1977; White \&
Rees 1978). Massive galaxies do however accrete large hot envelopes of
gas where the cooling time exceeds the infall time but is still less
than the age of the Universe; i.e. they have cooling flows (Nulsen \&
Fabian 1995). Whether such gas cools and builds galaxies further or is
heated and prevented from cooling is not known. Heating by a central
engine is one solution. The consequence of this is that the central
engine (the growth of the central black hole) then has a profound
influence on the final state of the galaxy.

Further observations of the soft X-ray phenomenon in cooling flows
therefore have potential implications for our understanding of cooling
and heating generally, as well as for the uniformity of metal
enrichment of hot gas, the energy content of jets, and galaxy
formation.

\section{Acknowledgements}
We thank Steve Kahn, Frits Paerels, Jelle Kaastra and Takayuki Tamura
for discussions on RGS spectra and Robert Schmidt and Steve Allen for
discussion on the Chandra data. ACF thanks the Royal Society for
support.

\end{document}